\documentclass[aps, pra, reprint, showpacs,preprintnumbers, a4paper, superscriptaddress]{revtex4-1}
\usepackage[utf8x]{inputenc}
\usepackage[english]{babel}
\usepackage{graphicx}
\usepackage{dcolumn}
\usepackage{amsmath}
\usepackage{amssymb}
\usepackage{placeins}
\usepackage{grffile}
\usepackage{bm}
\usepackage{color}        
\definecolor{darkred}{rgb}{0.5,0.0,0.0}

\begin{document}
\title{Stabilization of localized structures by inhomogeneous injection in Kerr resonators }
\author{Felix Tabbert}
\email{felix.tabbert@uni-muenster.de}
\affiliation{Institute for Theoretical Physics, University of M\"unster,
Wilhelm-Klemm-Str.\,9, D-48149 M\"unster, Germany}
\author{Tobias Frohoff-H\"ulsmann}
\affiliation{Institute for Theoretical Physics, University of M\"unster,
Wilhelm-Klemm-Str.\,9, D-48149 M\"unster, Germany}
\author{Krassimir Panajotov}
\affiliation{Vrije Universiteit Brussel, Department of Applied Physics and Photonics,
Pleinlaan 2, B-1050 Brussels, Belgium.}
\affiliation{Institute of Solid State Physics, Bulgarian Academy of Sciences, 72 Tzarigradsko Chaussee Blvd.,  1784 Sofia,  Bulgaria}
\author{Mustapha Tlidi}
\affiliation{Faculté des Sciences, Université Libre de Bruxelles, Campus Plaine, C.P. 231, Brussels B-1050, Belgium}
\author{Svetlana V. Gurevich}
\affiliation{Institute for Theoretical Physics, University of M\"unster,
Wilhelm-Klemm-Str.\,9, D-48149 M\"unster, Germany}
\affiliation{Center for Nonlinear Science (CeNoS), University of M\"unster,
Corrensstr.\,2, D-48149 M\"unster, Germany}
\date{\today}
\begin{abstract}

We consider the formation of temporal localized structures or Kerr comb generation in a microresonator with inhomogeneities. We show that the introduction of even a small inhomogeneity in the injected beam widens the stability region of localized solutions. The homoclinic snaking bifurcation associated with the formation of localized structures and clusters of them with decaying oscillatory tails is constructed. Furthermore, the inhomogeneity allows not only to control the position of localized solutions, but strongly affects their stability domains. In particular, a new stability domain  of  a  single peak localized structure appears outside of the region of  multistability between multiple peaks of localized states. We identify a regime of larger detuning, where localized structures do not exhibit a snaking behavior. In this regime, the effect of inhomogeneities on localized solutions is far more complex: they can act either attracting or repelling. We identify the pitchfork bifurcation responsible for this transition. Finally, we use a potential well approach to  determine the force exerted by the inhomogeneity and summarize with a full analysis of the parameter regime where localized structures and therefore Kerr comb generation exist and analyze how this regime changes in the presence of an inhomogeneity.
\end{abstract}
%
%
%
\maketitle
\section{Introduction}
The formation of localized structures (LSs) is a fascinating pattern formation phenomenon that has been experimentally observed and theoretically described in a wide variety of fields ranging from fluid mechanics, optics, chemistry, to plant ecology  \cite{pomeau1986pomeau,akhmediev2008dissipative, purwins2010dissipative, liehr2013dissipative,tlidi1994localized,leo2010temporal,umbanhowar1996localized, vanag2007localized, Tlidi20140101,meron2015nonlinear,Lugiato_book_2015,LLE_Specailissue,tlidi2018dissipative}.

In the field of nonlinear optics, LSs have been intensively studied theoretically and observed experimentally in both spatial and temporal domains. In particular, spatial LSs have been observed in the transverse section of broad-area semiconductor microcavities injected by a coherent electromagnetic field~\cite{barland2002cavity}.  More recently, the question whether the concept of LS can be extended to the time domain in the case of optically injected cavities~\cite{leo2010temporal} was addressed and experimentally observed. This behavior has been theoreticaly predicted in an early report \cite{scroggie1994pattern}.
A key role for the theoretical investigation of LSs in nonlinear optics plays the paradigmatic Lugiato-Lefever equation (LLE), a model first proposed by Lugiato and Lefever \cite{Lugiato_prl87} to describe spatial pattern formation in the transversal plane of a cavity filled with a nonlinear Kerr-medium. Later on it was shown, that the LLE also applies to the formation of temporal LSs in a ring cavity by replacing diffraction by group velocity dispersion \cite{haelterman1992dissipative}. The investigation of temporal LSs in the LLE has gained significant new importance in relation with the generation of optical
frequency comb generation. Kerr combs consist of a multitude of equidistant coherent spectral lines and directly link the optical and radio frequency band of the electromagnetic spectrum \cite{Hansch_rmp06}. The observation of broadband optical frequency combs has been realized in high finesse resonators filled with a Kerr medium and driven by a continuous wave \cite{DelHaye_nature07}. Frequency-combs generated in passive optical Kerr
resonators are in fact nothing but the spectral content of the temporal LS occurring in the
cavity. Indeed, the link between the LLE and the generation of optical frequency combs has been established in \cite{matsko2011mode}. Recently, an excellent overview by the Lugiato and Kippenberg groups has been published in which they discuss in depth the link between temporal LSs and optical frequency combs \cite{PTRS_Lugiato_2018}.

It has been shown analytically and experimentally that a focusing Kerr resonator driven by an inhomogeneous Gaussian pumping beam supports stable LSs \cite{Odent2014}. These structures result from front interaction in avregime devoid of modulational instability.  The
trajectory of the position of LS is derived from the LLE and its hyperbolic tangent analytical expression perfectly fits the experimental data  \cite{Odent2014}. 
In that case, the CW and the Gaussian beams are derived from the same pump laser. We suppose, that one can also derive a strong CW and the weakly modulated beams from the same laser for the case of temporal localized structures. Indeed, a synchronously pumped passive all fiber Kerr cavity (modulated single pump beam) has been realized in e.g. \cite{coen1999convection,xu2014experimental}.  Recently, Hendry et al \cite{HendryPRA2018} have considered a Gaussian pumping in the LLE. In particular, it was shown that LSs do not necessarily stabilize at minima or maxima of the injection but instead are drawn towards specific ideal values of the injection. Furthermore, recent work of Cole et al \cite{Cole_Optica_2018} suggests, that a phase-modulated injection can protect single LS generation by preventing the
multistability between different LSs having different number of peaks.

In this work we provide a systematic analysis of the impact of small inhomogeneities altering the amplitude of an otherwise homogeneous injected pumping on LSs dynamics. The consideration of small inhomogeneities seems inevitable, because it is difficult to prevent them in any real experimental setup. However, even small inhomogeneities can have drastic effects on the dynamical properties of a system under consideration because they break continuous symmetries of the system \cite{tabbert17}. It is therefore necessary to include these symmetry breaking effects in a theoretical description. Furthermore, we are going to demonstrate, that the addition of inhomogeneities can also be beneficial for Kerr comb generation. In certain scenarios it is therefore not necessary to minimize the inhomogeneities, but one can take advantage of them instead.

Employing path-continuation techniques, we start in a parameter regime where LSs arise in a homoclinic snaking \cite{GOMILA2007,tlidi2010high, Parra_Rivas_PRA_2014}. We are going to show how the inclusion of small inhomogeneities alters the snaking behavior by drastically widening the parameter regime in which stable LSs exist. Furthermore, a parameter gap arises, in which only a single LS positioned at the inhomogeneity is stable, thus avoiding the multistability associated with the homoclinic snaking. Both results suggest that small inhomogeneities can actually be beneficial for the experimental realization of  Kerr comb generation.

We then proceed with a similar analysis in a regime of higher detuning, where the results from \cite{HendryPRA2018} come to fruition. Since in this regime LSs are drawn towards certain specific values of the injection, the bifurcation structure becomes much more complex. We identify three different stationary solutions in this regime: LSs can be (a) pinned on the center of the inhomogeneity, (b) pinned on the side of the inhomogeneity or (c) can be completely repelled by the inhomogeneity. We describe all these scenarios and the transition between them. Further, we deploy a semi-analytic potential well model that allows to determine the position of a single LS. Finally, we provide a full description of the region of existence and the region of stability of a single LS in the inhomogeneous LLE in terms of the two main control parameters, the detuning and the injection. This result is a full bifurcation diagram, showing where Kerr combs generation is theoretically possible.

\section{The Model}

The starting point of this study is the generic dimensionless focusing mean-field LLE with inhomogeneous injection that reads:
\begin{equation}
\frac{\partial E}{\partial t}=E_\text{inh}(\xi)+\left[-(1+i\theta) +i|E|^2+i \frac{\partial^{2}}{\partial \xi^{2}}\right]E.
\label{LLE}
\end{equation}
Here, the intracavity field envelope is denoted by $E=E(t,\xi)$, $\theta$ is the detuning parameter. In the context of temporal LSs in a ring-cavity, $\xi$ is the fast time in the reference frame moving with the group velocity of the light within the cavity while $t$ is the slow time proportional to the round-trip time. In the originally proposed LLE describing spatial pattern formation, $\xi$ is the spatial coordinate in the transversal plane of a cavity and $t$ is the time. In that case, an inhomogeneous injection $E_\text{inh}(\xi)$ with the CW and the Gaussian beams derived from the same pump laser has been realized in a resonator with a liquid crystal as Kerr media \cite{Odent2014}. We suppose, that one can also derive a strong CW and the weakly modulated beams from the same laser for the case of temporal localized structures. Indeed, a synchronously pumped passive all-fiber Kerr cavity (modulated single pump beam) has been realized in e.g. \cite{coen1999convection,xu2014experimental}. For the sake of simplicity we refer in both scenarios to the inhomogeneity as spatial inhomogeneity.  The inhomogeneous injected beam $E_\text{inh}(\xi)$ reads

\begin{equation}
E_\text{inh}(\xi)=E_{i}+A~\text{exp}(-\xi^2/B),
\end{equation}
where $E_{i}$ is the homogeneous value of the injection, $A$ and $\sqrt{B}$ correspond to the amplitude and the width of the Gaussian beam, respectively.  However, our results suggest that the overall influence of inhomogeneities on LSs mainly depends on the amplitude of the inhomogeneity $A$. Neither the exact form nor the width $\sqrt{B}$ of the inhomogeneity have an equally important effect within a reasonable range. That is, as long as the width of the inhomogeneity is smaller than the considered domain size $L=100$ and of comparable size as the typical length-scales in the system (e.g. the width of the LS or the wave-length of the periodic patterns), varying $B$ does not change the solution structure qualitatively. Hence, we will focus on the influence of different values of $A$ in the following, while leaving the width of the inhomogeneity fixed at $\sqrt{B}=2.0$ and its Gaussian shape remains unaltered. 

In the case of homogeneous injection ($A=0$) Eq.\eqref{LLE} represents the original LLE as proposed by Lugiato and Lefever \cite{Lugiato_prl87}. The case of purely Gaussian injection ($E_i=0$) has been recently discussed by Hendry et al \cite{HendryPRA2018}. We are going to focus on the case of a homogeneous injection $E_i$ with a small added inhomogeneity and mainly discuss how these inhomogeneities alter the well-known properties of the classical LLE with purely homogeneous injection. Nevertheless, we are going to demonstrate in section \ref{Sec:Nosnake} that the essential result of \cite{HendryPRA2018} also applies to the scenario of small inhomogeneities.

The classical LLE with homogeneous injection $A=0$ has been thoroughly studied~\cite{Lugiato_prl87, scroggie1994pattern, GOMILA2007, Parra-Rivas2017_Interaction}. Homogeneous stationary solutions $E_s$ of Eq. \eqref{LLE} are implicitly given by $E_{i}^2=|E_s|^2[1+(\theta -|E_s|^2)^2]$. For $\theta <\sqrt{3}$
($\theta >\sqrt{3}$) the transmitted intensity $|E_s|^2$ as a function of the input intensity $E_{i}^2$ is monostable (bistable) \cite{scroggie1994pattern}. Localized solutions exist in both regimes \cite{Parra_Rivas_PRA_2014}. The homogeneous solution looses stability in a modulational (Turing like) instability that is subcritical (supercritical) for $\theta>41/30$ ($\theta<41/30$). A necessary condition for the existence of LSs is a bistability between a homogeneous and a periodic solution, which is only given in the subcritical case. In this case, the periodic solution first branches off of the homogeneous solution at the Turing bifurcation point and is originally unstable, and then gains stability in a fold. Without inhomogeneities, LSs bifurcate at the same point as the periodic solution and become stable after a fold.

In the regime where LSs possess oscillatory tails ($\theta\lesssim 2$), bound states can form in a so-called homoclinic snaking~\cite{GOMILA2007,tlidi2010high, Parra_Rivas_PRA_2014, Parra-Rivas2017_Interaction}: In a sequence of consecutive folds, the LS solution gains two peaks until the solution fills the domain and connects to the periodic solution. Besides the branch always containing an odd number of peaks, there also exists an even branch starting from a bound solution of two peaks. The odd branch in the case of homogeneous injection is depicted in Fig. \ref{fig:theta1.7} (blue line). The homoclinic snaking structure in the classical LLE has been intensively studied \cite{GOMILA2007,tlidi2010high,Parra_Rivas_PRA_2014}, however, it is a general phenomenon that can be found in a number of systems possessing LS solutions \cite{champneys1998homoclinic,hunt1999homoclinic,coullet2000stable} (see overviews on this issue \cite{BurkeKnoblochChaos2007,knobloch2008spatially}). So far, however, the impact of defects or inhomogeneities on the snaking bifurcation structure has not been discussed yet, and we will discuss it in section \ref{Sec:Snaking}.

In the regime $\theta\gtrsim 2$, LSs can not form bound states due to a lack of oscillatory tails, i.e. two LSs always act repulsive on each other. Homoclinic snaking can therefore not be observed. The origin of a single LS however remains unchanged and we will discuss the influence of inhomogeneities on this solution in section \ref{Sec:Nosnake}. For even higher values of the detuning $\theta$ a single LS becomes unstable in an Andronov-Hopf bifurcation and starts to oscillate~\cite{Parra_Rivas_PRA_2014}. We are going to consider this effect in the last section and show how the position of the Hopf bifurcation is affected by inhomogeneities.

\section{Homoclinic Snaking in the Presence of Spatial Inhomogeneities}\label{Sec:Snaking}
\begin{figure}
\hspace*{-0.35cm}
\includegraphics[width=0.5\textwidth]{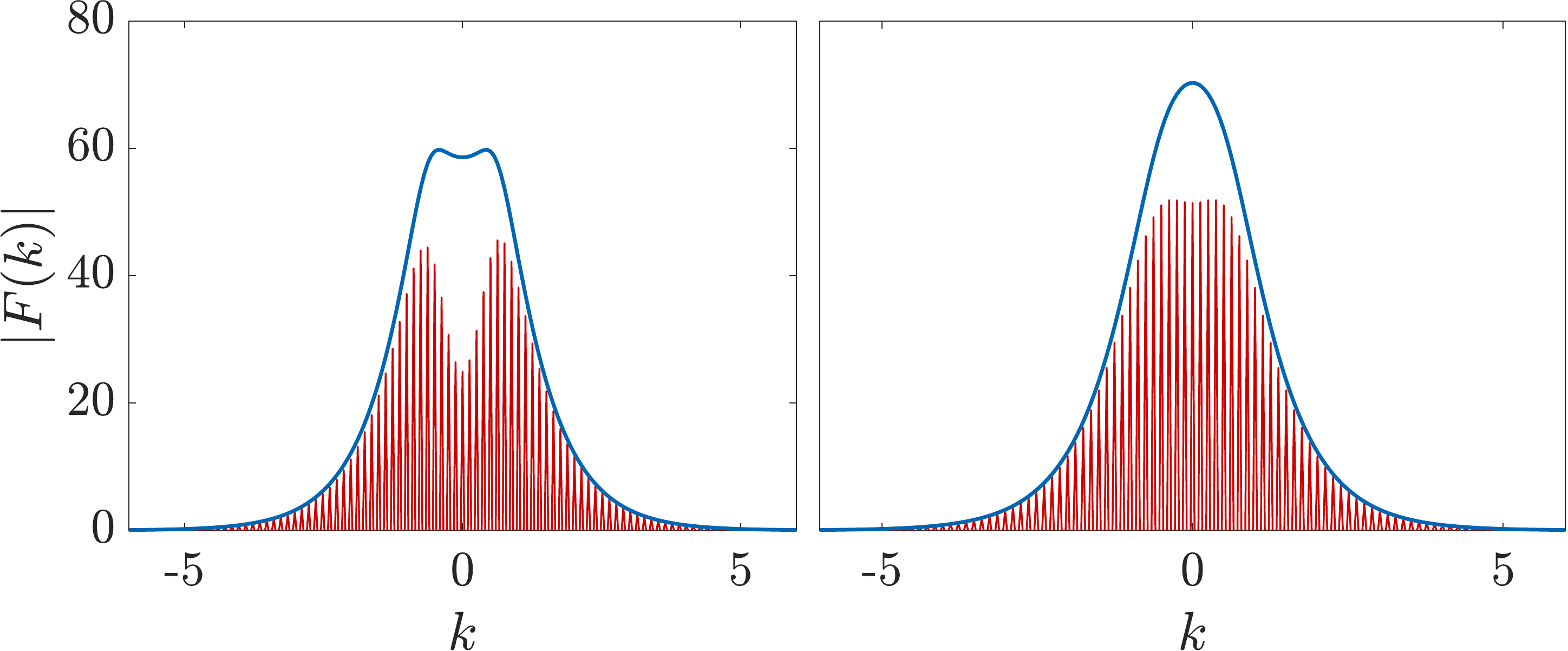}
 \caption{Kerr combs generated by Fourier transforming a single LS of Eq. \eqref{LLE}. The blue line indicates the absolute value of the Fourier transform, whereas the red line indicates the real value. On the left, the Kerr comb with homogeneous injection ($A=0$) is obtained for $\theta=1.7$ and $E_i=1.2$. On the right, frequency comb with a small inhomogeneity ($A=0.1$, $B=4.0$) for the same value of $\theta$ but with a lowered value of $E_i=1.12$ is shown.}
 \label{fig:OFC}
\end{figure}

In this section we are focusing on the homoclinic snaking regime of Eq. \eqref{LLE} and fix the detuning value to $\theta=1.7$. To demonstrate that even with small inhomogeneities, temporal LSs of the LLE can serve as a useful source for Kerr combs, Fig. \ref{fig:OFC} shows the frequency comb generated by a LS in the classical LLE (left) and in the presence of a small inhomogeneity ($A=0.1$) (right panel). This comparison shows that one is not only able to generate Kerr combs in the presence of inhomogeneities, but one also needs in less injected energy to create a comparable comb, since the homogeneous portion of the injection $E_i$ has been lowered on the right of Fig. \ref{fig:OFC}. Since the connection of comb generation  with LSs of the LLE is well established~\cite{PTRS_Lugiato_2018}, we will from here on focus on the properties of localized solutions.

\begin{figure}
\hspace*{-0.35cm}
\includegraphics[width=0.5\textwidth]{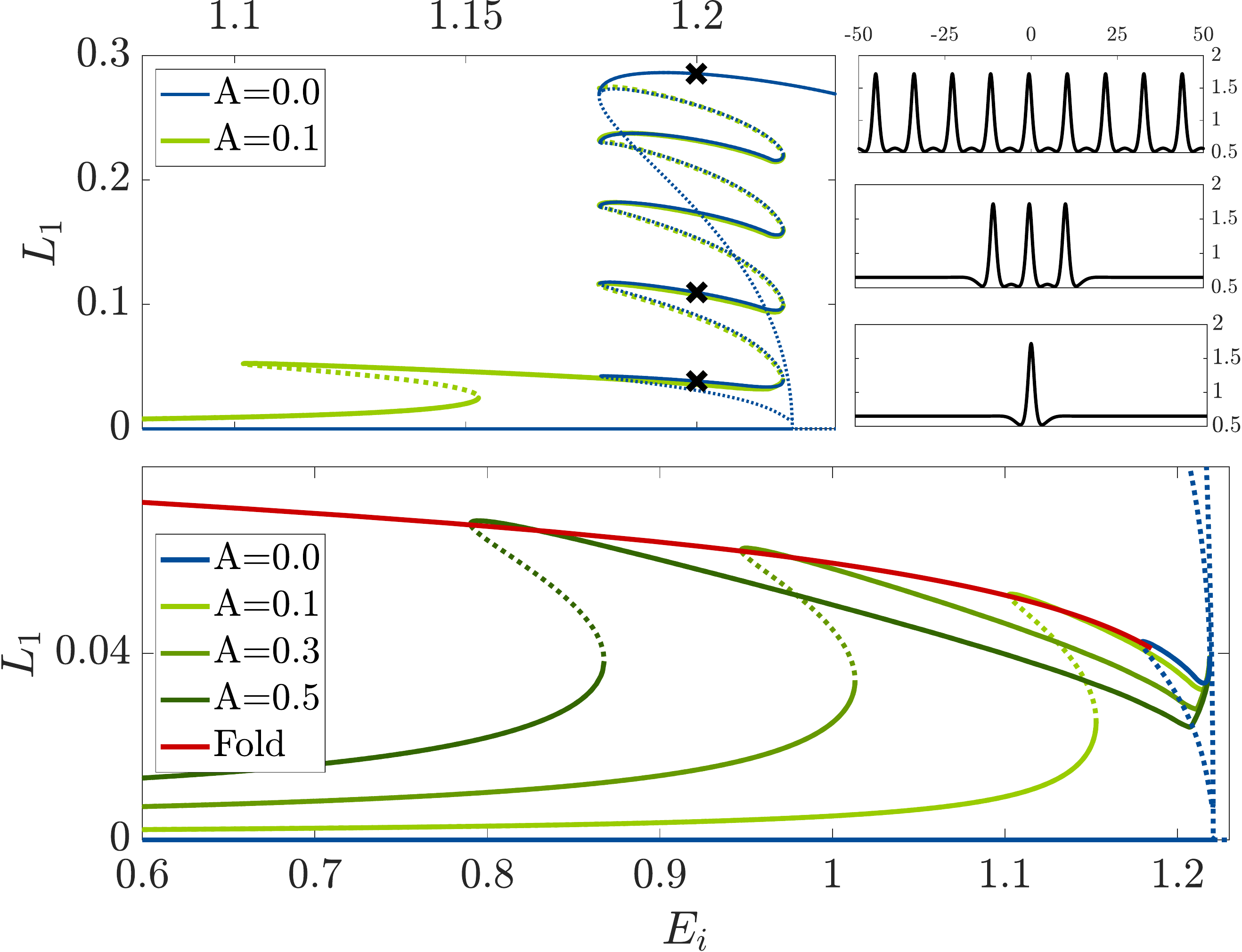}
 \caption{Top: (left) Bifurcation snaking diagram with and without inhomogeneity for $\theta=1.7$. The blue line depicts the classical snaking without inhomogeneity($A=0$), whereas the green line shows the homoclinic snaking in the presence of a small inhomogeneity ($A=0.1$). Black crosses show the position of three exemplary solutions, whose real parts $E_r(\xi)$ are depicted in the right panels.
 Bottom: Different sector of the upper diagram depicting the evolution of the left fold point with increasing inhomogeneity $A$ (green branches). The red curve continuously marks the position of the fold point while altering $A$.}
 \label{fig:theta1.7}
\end{figure}

The LS c in Fig. \ref{fig:OFC} is positioned directly on the inhomogeneity, i.e. in this case the inhomogeneity acts attracting on the LS. To understand the effect of the inhomogeneities on the bifurcation structure, we deploy numerical continuation techniques provided by the Matlab continuation package \textit{pde2path} \cite{uecker_wetzel_rademacher_2014}. In Fig. \ref{fig:theta1.7}, we plot the $L_1$ norm $L_1=\int d\xi|\text{Re}(E-\overline{E})|$ as a function of $E_i$ for different solutions of the LLE.  $\overline{E}$ denotes the mean value of the electrical field $E(\xi)$ averaged over the domain size. We chose this definition of a norm, since the real part of the LSs is more pronounced than the imaginary part, which makes it easier to differentiate different solution branches. One could however, chose a different solution measure that allows to distinguish different solution branches in a bifurcation diagram. The blue line shows the odd branch of the classical homoclinic snaking in the case of homogeneous injection ($A=0$).  The homogeneous solution ($L_1=0$) looses stability at the Turing point, where both the periodic as well as the single peak localized solution bifurcate subcritically. The same goes for the even snaking branch initially consisting of a bound state of the LS, however, we abstain from including this branch in Fig. \ref{fig:theta1.7} for the sake of clarity. The single peak solution reaches stability in a fold and then gains additional peaks in a sequence of folds until the domain is filled. Solution profiles during this snaking process at the position marked in the bifurcation diagram are depicted in Fig. \ref{fig:theta1.7} on the upper-right panel.  The impact of adding extra peaks is manifested in an extra modulation of the frequency comb and the modulation depth becomes more pronounced with the number of the peaks as shown numerically in \cite{Parra_Rivas_PRA_2014}.

The green line in the upper left of Fig. \ref{fig:theta1.7} shows the bifurcation diagram in the presence of a small inhomogeneity of $A=0.1$. Starting at the far left with a quasi-homogeneous solution (that is, a homogeneous solution with a slight deformation at the position of the inhomogeneity), this solution looses stability and transforms into a single peak LS in a fold, which then becomes stable in another fold. From there on, the rest of the bifurcation diagram is hardly affected by the inhomogeneity. Focussing on the region of stability of the single peak LS, it is not surprising, that the left fold demarking the onset of stability of the LS shifts to the left compared to the classical LLE, because this solely means, that smaller amounts of overall homogeneous injection $E_i$ are needed when there is an additional positive inhomogeneous injection. However, it is more surprising, that the position of the right fold, limiting the stability of the single LS, is hardly affected  by the inhomogeneity. This phenomenon provides valuable insights in the formation of LSs: Whereas the injection at the peak position seems to determine the onset of existence, the existence delimiting factor going to large injection intensities seems to be the total injection value at the sides of the LS. The shift of the left fold position results in a drastic enlargement of the region of stability of single LS . Furthermore a region emerges, where solely the single LS solution is stable, avoiding the multistability between a single LS, the homogeneous solution and LSs consisting of more than one peak that exists in the case of the classical homoclinic snaking without inhomogeneity. One can argue whether or not a single peak solution without a stable background still can be classified as a LS, however we choose to do so in the following because there is no qualitative difference in the solution profile between the region where the LS coexists with a quasi-homogeneous background and the region where only the LS exists as a stationary solution. Regardless of the nomenclature, experimental work in this parameter regime can drastically simplify Kerr combs generation because one can easily address the single peak solution avoiding unwanted jumps to other solutions which can occur in regions of multistability.

Figure \ref{fig:theta1.7} on the bottom shows, that one can increase this favorable parameter regime by using larger inhomogeneities. The regions of stability for single LS are depicted in shades of green for increasing values of $A$, showing a growing region of monostability. For further analysis it is useful to calculate the onset of single LS stability as a function of $A$. We therefore deploy numerical fold-point continuation to track the position of the left fold. To do so, we choose $A$ as our main continuation parameter and choose $E_i$ as a free parameter which is determined by additional conditions characterizing the fold. The red line in Fig. \ref{fig:theta1.7} depicts the result of this fold-point continuation, showing that one can further increase the region of monostability. In fact, choosing an inhomogeneity of $A=0.5$, one can increase the region of stability by more than order of magnitude compared to the case of homogeneous injection.
\begin{figure}
\hspace*{-0.35cm}
\includegraphics[width=0.5\textwidth]{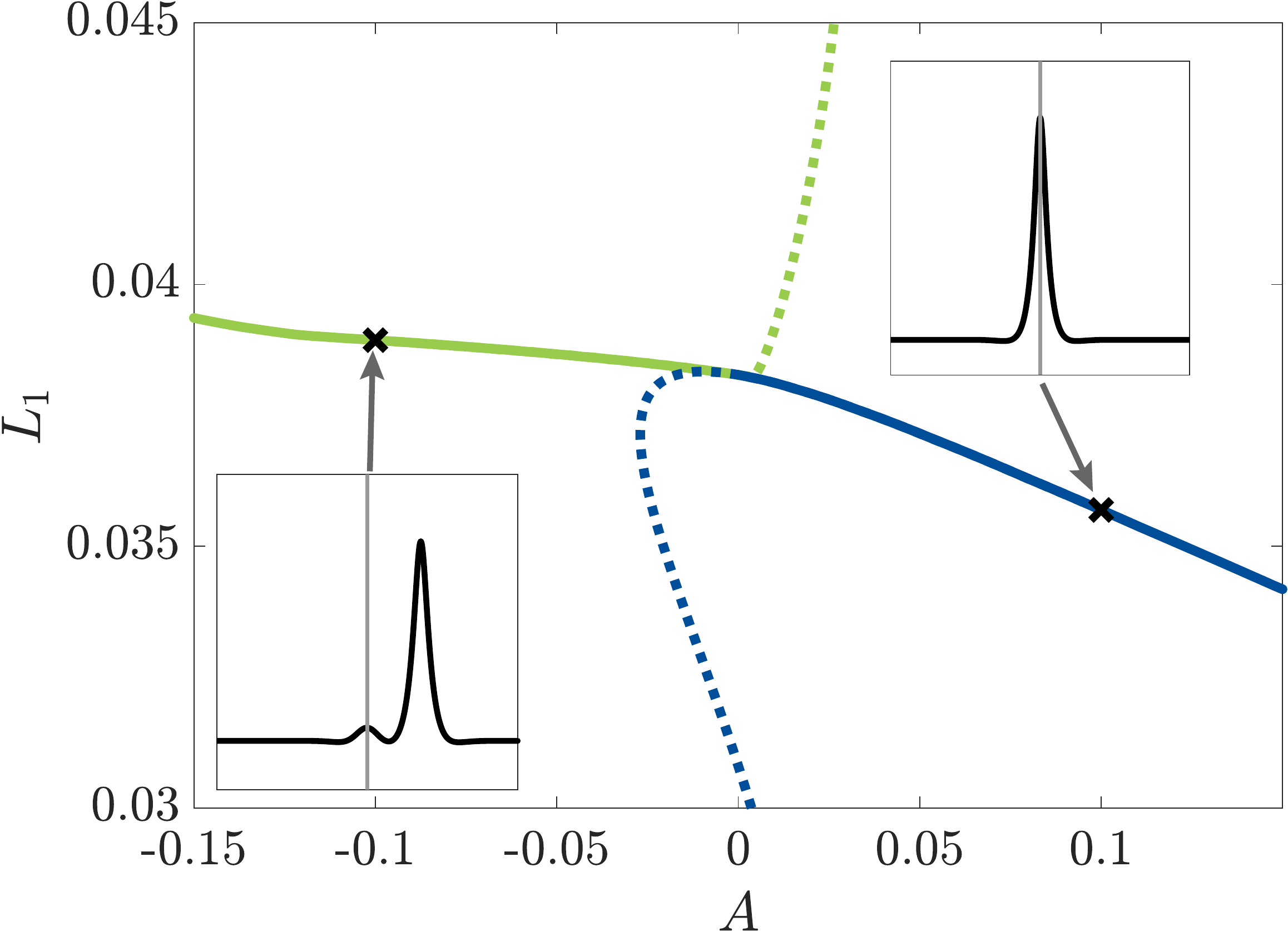}
 \caption{Bifurcation diagram for $\theta=1.7$ and $E_i=2.0$, with $A$ as the main continuation parameter. The blue line corresponds to a LS pinned on the center of the inhomogeneity, which is stable for positive values of $A$ and interchanges stability with a solution pinned on the side of the inhomogeneity (green line) at $A=0.0$. The insets depict the solution profiles of Re$(E)$ at the marked positions.}
 \label{fig:Atheta1.7}
\end{figure}

In order to demonstrate, how different amplitudes of inhomogeneities affect the stable and unstable solutions of the system in question, we also perform numerical continuation with $A$ as the main continuation parameter. Results are depicted in Fig. \ref{fig:Atheta1.7}. As already discussed above, for positive $A$ the inhomogeneity acts attracting on LSs and therefore a stable solution pinned on the center (blue line) exists. At $A=0.0$ this solution interchanges stability in a transcritical bifurcation with a solution pinned on the side of the inhomogeneity (green line). Note that, although the solutions are situated at different positions at all time, we can still identify this transition as a transcritical bifurcation since in the case of $A=0.0$, both solutions are mathematically identical due to the translational symmetry realized by periodic boundaries we are assuming for the numerical continuation. The effects of the inhomogeneities shown in Fig. \ref{fig:Atheta1.7} are fairly intuitive. In the next section we are going to use similar techniques, demonstrating however, that the influence of inhomogeneities in regions of higher values of the detuning $\theta$ can be much more complex.

\section{The non-snaking Regime}\label{Sec:Nosnake}
\begin{figure}
\hspace*{-0.35cm}
\includegraphics[width=0.5\textwidth]{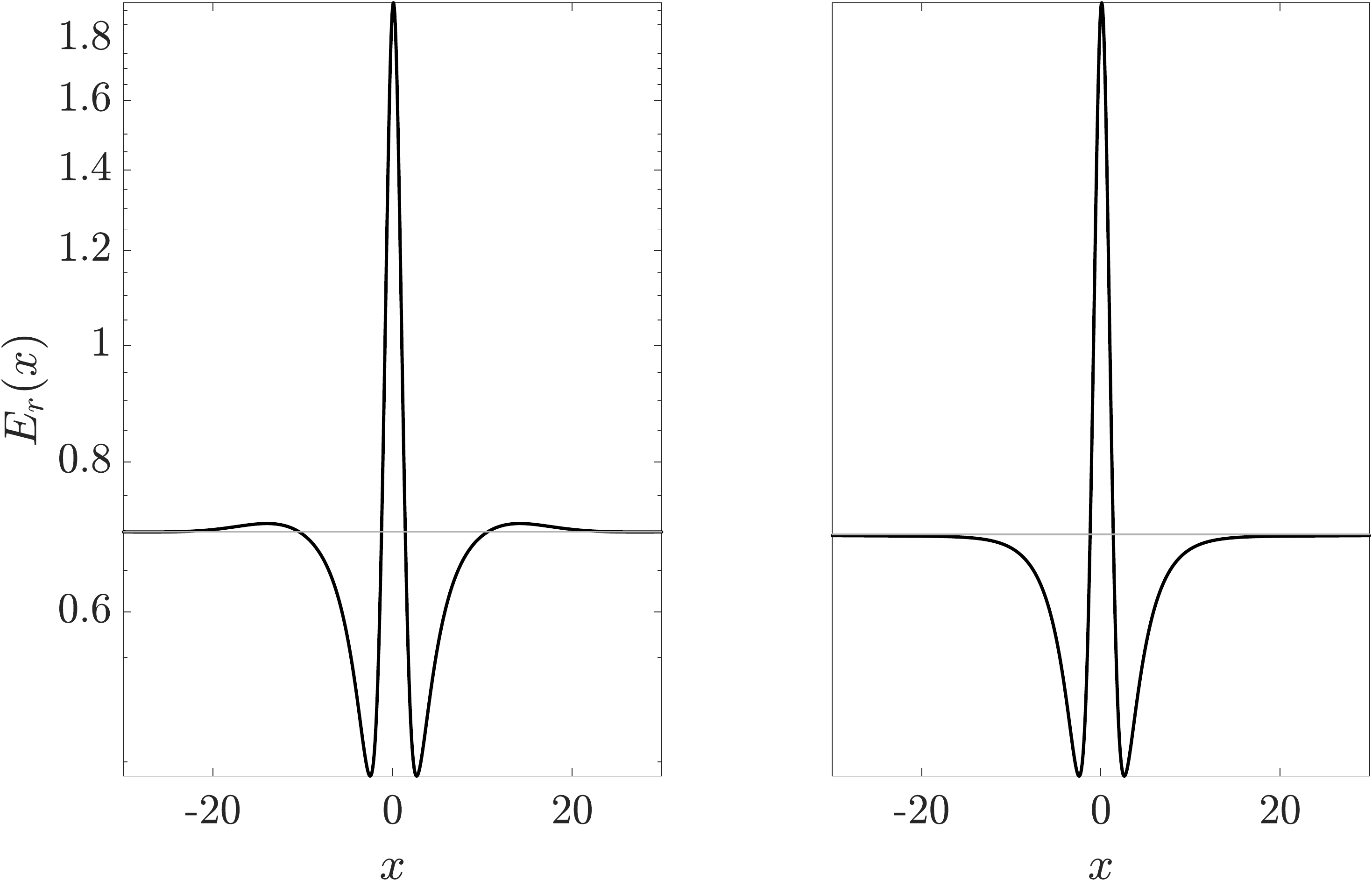}
 \caption{Solution profile of a LS for $\theta=2.0$ (left) and $\theta=2.1$ (right) at the right fold delimiting the stability regime of the LS. The grey horizontal line and the logarithmic scale emphasize the transition from a localized solution with an additional maximum on each side to a localized solution where these maxima have vanished. In the regime $\theta>2.085$ where these maxima are abscent, a bound state between two LSs is not possible, the two LSs act repelling}
 \label{fig:tails}
\end{figure}

In this section, we are going to focus on larger values of the detuning $\theta$. In \cite{Parra_Rivas_PRA_2014}, by applying a linear stability analysis it was shown, that even for large values of $\theta$, LSs possess oscillatory tails, although they are less pronounced than in the case of lower detuning. Oscillatory tails represent a necessary condition for the stability of bound states and therefore represent a necessary condition for the occurrence of homoclinic snaking. However one can see that for $\theta>2.0$, the oscillatory tail becomes less and less pronounced. As shown in Fig. \ref{fig:tails}, the stable single LS for $\theta=2.0$ (left) possesses one side-minimum and a small side-maximum, which however vanishes for $\theta=2.1$ (right). The LS solution still has an oscillatory tail, the latter however gets nonlinearly suppressed so that only one side-extremum (the minimum) remains. Yet the vanishing maximum is necessary for the formation of bound states and therefore the snaking structure vanishes at $\theta_{crit}\approx 2.085$. For larger values of $\theta$, two LSs always act repelling, however the repelling effect is rather weak and therefore hard to detect in e.g. direct numerical simulations of Eq. \eqref{LLE}. Above the value of $\theta_{crit}$, the emergence of a single LS remains qualitatively the same as in Fig. \ref{fig:theta1.7}, however, after the single LS looses stability, additional peaks do not form at the side of the existing LS. Instead an additional peak arises at the maximum possible distance to the existing LS. Branches consisting of several solutions can exist, depending on the domain size and the precision of the used continuation algorithm, yet they are not connected to the branches with lower LSs number.

When considering the influence of inhomogeneities in this parameter regime, one also has to take into account the recent results of Hendry et al \cite{HendryPRA2018} who have shown that for larger values of $\theta$ a LS is not necessarily drawn towards the maximum of the injection field, but that there exists a certain ideal value of injection depending on $\theta$, that act attracting on LS. In contrast to \cite{HendryPRA2018} we are going to consider small inhomogeneities, however we are going to demonstrate that the described effect is also important for the present work.

\begin{figure}
\hspace*{-0.35cm}
\includegraphics[width=0.5\textwidth]{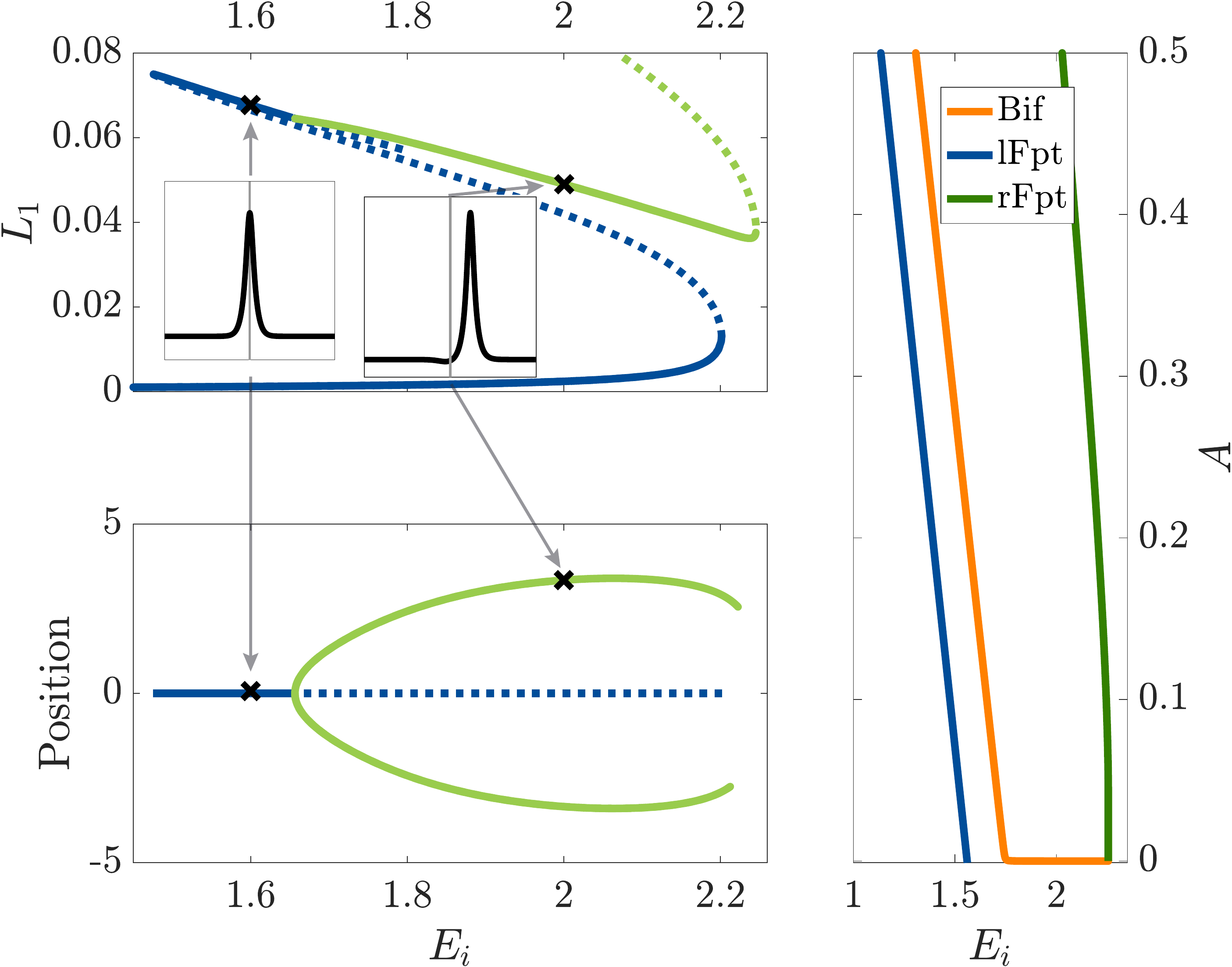}
 \caption{Upper Left: Bifurcation diagram in $(E_i, L_1)$ plane showing the emergence of LS for $\theta=3.0$ and $A=0.1$. A LS gains stability after two consecutive folds (blue line). However, the stable solution pinned on the center (left inset) of the inhomogeneity looses its stability in a pitchfork bifurcation at $E_i\approx 1.66$ giving rise to a new stable solution that is pinned on the side of the inhomogeneity (right inset, green line).
 Lower Left: Same bifurcation scenario with the position of the center of mass on the y-axis, clearly showing the transition from a stable solution pinned on the center to two different stable solutions pinned on either side of the inhomogeneity in a pitchfork bifurcation.
 Right:Bifurcation diagram in ($A$, $E_i$) plane showing the region of the stability of a single LS. The blue line corresponds to the position of the left fold, orange line shows the position of the pitchfork bifurcation and the green line marks the right fold in which solutions pinned on the side loose stability. Stable LSs pinned on the center are located between the blue and the orange line, whereas stable LSs pinned on the side can be found between the orange and the green line, respectively.}
 \label{fig:theta3}
\end{figure}

Figure \ref{fig:theta3} (upper left) depicts the emergence of a single LS for $\theta=3$ with a small inhomogeneity of $A=0.1$. As in the previous case, a LS bifurcates from the homogeneous solution and gains stability in a fold at $E_i\approx1.475$. The LS positioned at the center of the inhomogeneity (see Fig. \ref{fig:theta3}, left inset) becomes unstable in a pitchfork bifurcation taking place at $E_i\approx1.657$, where two different stable solutions that are positioned on the side of the inhomogeneity (Fig. \ref{fig:theta3}, right inset) emerge. On the lower left, the same bifurcation diagram using the center of mass position of the LS instead of the $L_1$-norm is shown, thus underlining the pitchfork character of the bifurcation by showing that indeed two new stable solutions (one on the left, one on the right of the inhomogeneity) branch off. Since both solutions are identical except for their position, they are indistingiushable in the upper-left representation.

This result is rather striking since it shows that a given inhomogeneity of $A=0.1$ can act either attracting or repelling depending on the homogeneous injection $E_i$, however it can be explained by considering the results of \cite{HendryPRA2018}: At the bifurcation the overall injection at the center of the inhomogeneity $E_\text{inh.}=E_i+A$ reaches the ideal value (in this case $E_\text{ideal}\approx1.75$). For larger values of $E_i$, the solution therefore is pulled towards the position, where the ideal value is present, leading to a shift of the stable structure with increasing $E_i$ as can be seen in the lower left of Fig. \ref{fig:theta3}. This drift in parameter space comes to a halt when the homogeneous portion of the injection $E_i$ reaches the ideal value, in that case the LS pins on the side of the inhomogeneity (i.e. with its first minimum on the center of the inhomogeneity).

On the right panel of Fig. \ref{fig:theta3} we show how the positions of the stability-delimiting folds and the pitchfork bifurcation change with increasing $A$. In contrast to the results in the previous section, where the parameter regime of stable LS drastically widened with increasing $A$, in this case the range of stability of LS (pinned on the center or on the side) shifts drastically to the left and broadens only slightly.

\begin{figure}
\hspace*{-0.35cm}
\includegraphics[width=0.5\textwidth]{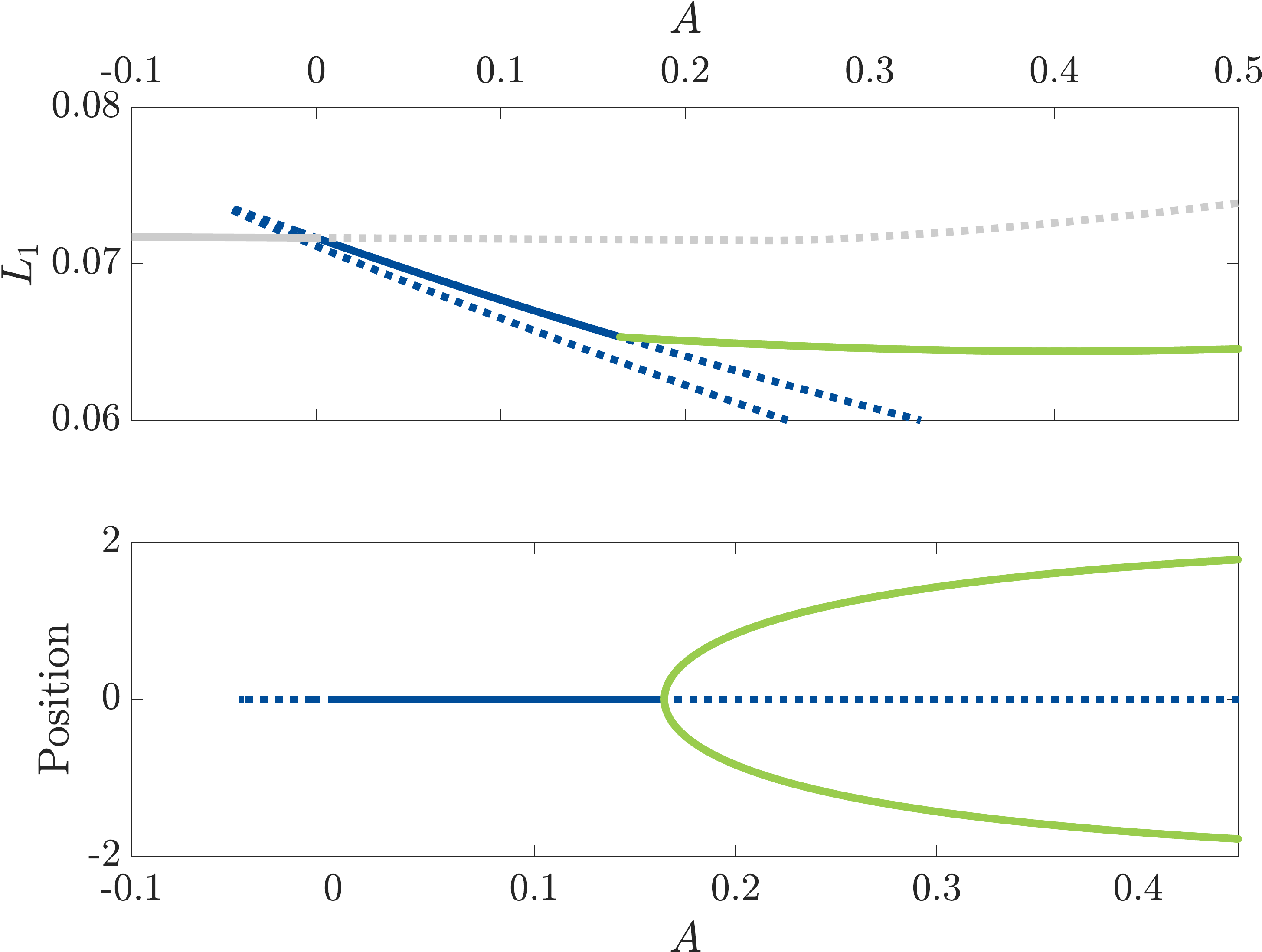}
 \caption{Bifurcation diagram for $\theta=3.0$, $E_i=1.6$ and varying $A$ with the $L1$-norm (top) and the center of mass position (bottom) as a measure. For small positive values of $A$ where $E_i+A$ is still below $E_\text{ideal}$, LSs pinned on the center of the inhomogeneity are stable (blue line), i.e. LSs are drawn towards the maximum value of injection. At the ideal value, two stable LSs pinned on either side of the inhomogeneity (green line) emerges in a pitchfork bifurcation. At $A=0$, the solution pinned on the center looses stability in a transcritical bifurcation interchanging stability with a solution positioned at the maximum possible distance to the inhomogeneity (gray line).}
 \label{fig:theta3E1.6}
\end{figure}

\begin{figure}
\hspace*{-0.35cm}
\includegraphics[width=0.5\textwidth]{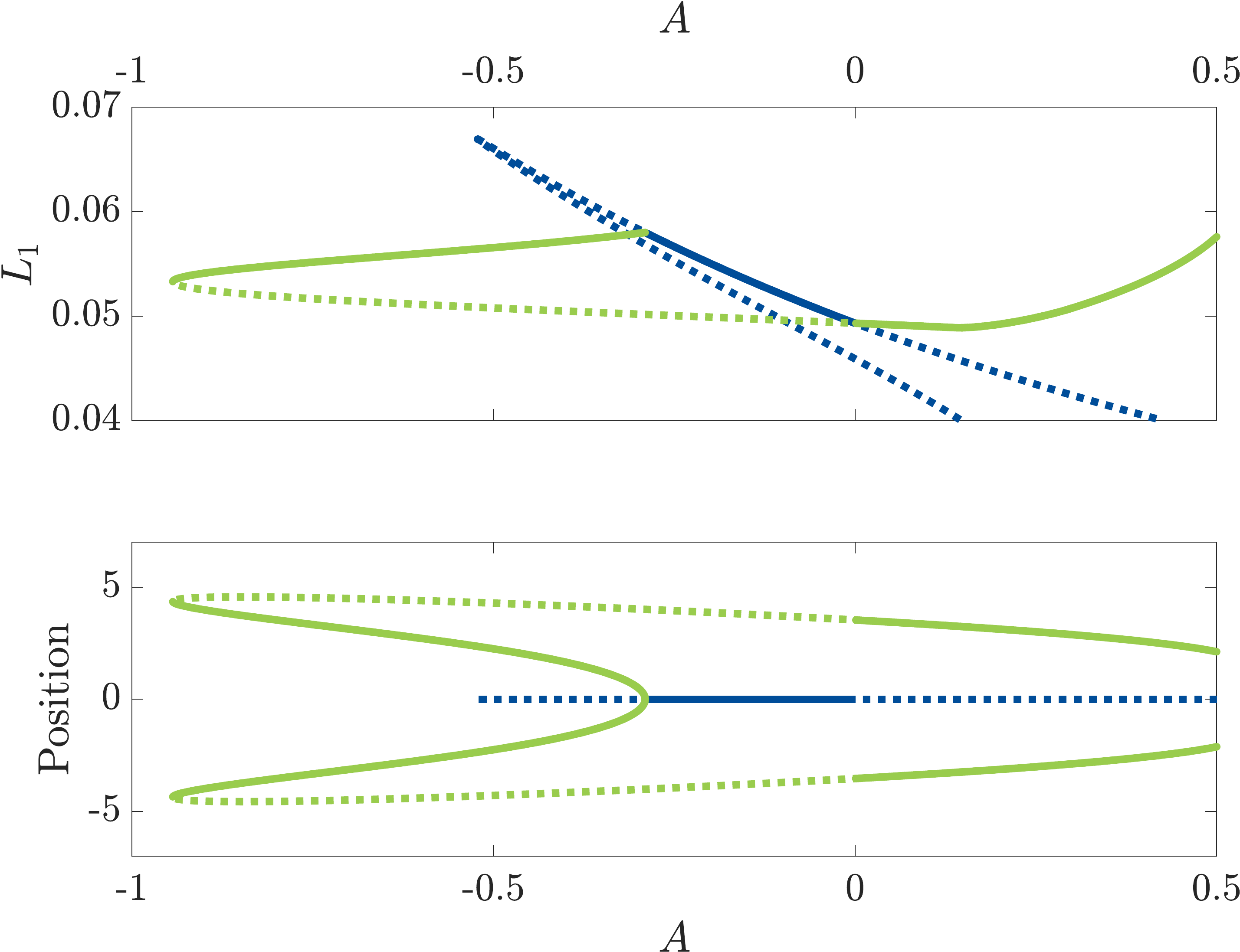}
 \caption{Bifurcation diagram for $\theta=3.0$, $E_i=2.0$ and varying $A$ with the $L1$-norm (top) and the center of mass position (bottom) as a measure. The solution pinned on the center (blue line) is stable for small negative values of $A$ until the overall injection at the center $E_i+A$ falls below the ideal value of injection $E_\text{ideal}$. At this point, a stable solution pinned on the side (green line) branches off before loosing stability in a fold. It is this solution that gains stability again in the transcritical bifurcation at $A=0.0$, where the solution pinned on the center looses its stability.}
 \label{fig:theta3E2.0}
\end{figure}

As in the previous section, we are now going to systematically analyze the influence of the amplitude $A$ of the inhomogeneity on LSs. As suggested by Fig. \ref{fig:theta3}, there are two fundamentally different regimes to perform this analysis: Figure \ref{fig:theta3E1.6} shows the bifurcation diagram of the system with $A$ as a main control parameter in the case where the homogeneous portion of the injection $E_i=1.6$ is still below the ideal value $E_\text{ideal}$, whereas Fig. \ref{fig:theta3E2.0} provides the same analysis in the case of $E_i=2.0$, where the homogeneous portion of the injection alone exceeds $E_\text{ideal}$. In both cases the region of stability of the LS on the center of the inhomogeneity is delimited by two bifurcations: A transcritical bifurcation at $A=0.0$ and a pitchfork bifurcation where the ideal value of injection at the center of the inhomogeneity is reached, i.e. where $E_i+A=E_\text{ideal}$.

In the case of $E_i=1.6$ at $A=0.0$ (cf. Fig. \ref{fig:theta3E1.6}) the system is still below the ideal value $E_\text{ideal}$, i.e. negative values of $A$ act repelling and the system only possesses a stable solution at the maximum distance to the inhomogeneity (gray line). The depicted stability change of this solution at $A=0$ is not obtained by numerical continuation techniques since the distance to the inhomogeneity of this solution is too large to detect the stabilizing or destabilizing influence of the inhomogeneity numerically. However, one can estimate the asymptotically vanishing influence of the inhomogeneity by analyzing the dynamics in direct numerical simulations or by applying the potential well model described in the next section. Note that no solutions pinned on the side of the inhomogeneity exist in this regime, since the tail of the LS in this case can not pin to the inhomogeneity with its first minimum.

Small positive values of $A$ act attracting on the LS, as long as the total injection $E_i+A$ at the center of the inhomogeneity is still below the ideal value $E_\text{ideal}$. If the total injection exceeds this value, the LS gets repelled from the center and is pinned at the position with the ideal injection, thus moving further away from the center of the inhomogeneity with increasing $A$, which can be seen in the lower panel of Fig.\ref{fig:theta3E1.6}.

For $E_i=2.0$, the situation is quite similar (cf. Fig. \ref{fig:theta3E2.0}): The centered solution is again stable between the ideal value of the injection and $A=0.0$, with the only difference being that the ideal value is now reached at a negative value of $A$. In other words: As long as the injection is above the ideal value in the whole domain, the LS moves towards the minimum of the injection at the center of the inhomogeneity. When the injection at the center falls below the ideal value, the pitchfork bifurcation sets in and the LS move towards the ideal value. In contrast to the case depicted in Fig. \ref{fig:theta3E1.6}, the solution pinned on the side undergoes a saddle-node bifurcation at large negative values of $A$ and then coexists as an unstable solution with the stable solution pinned on the center. At $A=0.0$ it is this solution pinned on the side that interchanges stability with the solution pinned on the center in a transcritical bifurcation. The transcritical bifurcation is clearly visibile in the upper representation of Fig. \ref{fig:theta3E2.0} at $A=0.0$. The lower representation, using the center of mass position as a measure, can be missleading, since the two lines interchanging stability do not cross. However, due to the translational symmetry, which is restored at $A=0.0$, both solutions are mathematically identical even though they differ in the center of mass position.

\section{Potential Well Model}
\begin{figure}
\hspace*{-0.35cm}
\includegraphics[width=0.5\textwidth]{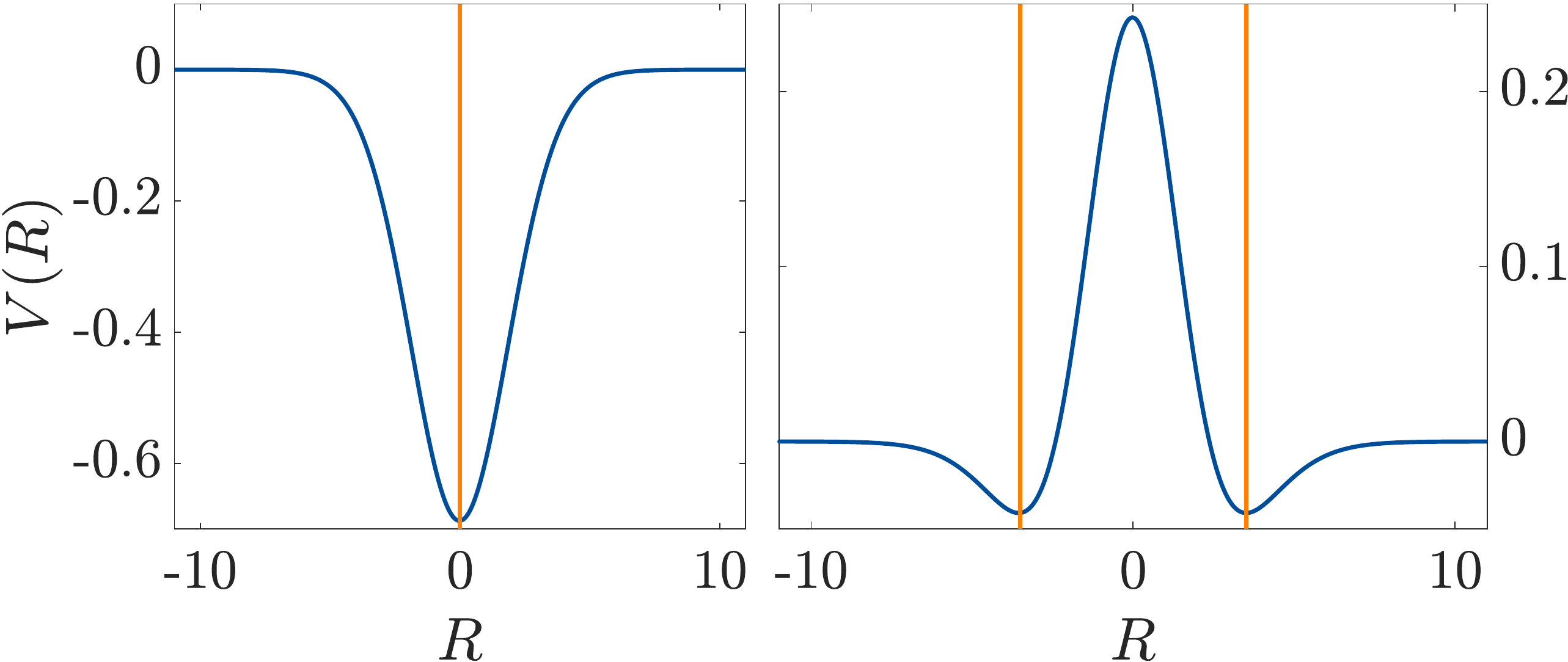}
 \caption{Potential $V(R)$ defined by Eq.~\eqref{pot} and calculated for $\theta=3.0$, a constant inhomogeneity of $A=0.1$ and two different values of the injection. In accordance with the results from Fig. \ref{fig:theta3}, the inhomogeneity acts attracting for $E_i=1.6$ (left) and repelling for $E_i=2.0$. The orange lines mark the maximum position of stable LSs found for the given parameters obtained by direct numerical simulations, showing that the solutions pin to the minima of the potential.}
 \label{fig:pot}
\end{figure}

As established in the previous section, a given inhomogeneity of fixed $A$ can act either attracting or repelling depending on the amount of overall injection. To further analyze the transition from an attracting to a repelling inhomogeneity, we deploy a semi-analytical method that consists of considering the LS in the vicinity of the inhomogeneity as an overdamped particle in a potential well. To calculate the force exerted by the inhomogeneity (at position $R=0$) on a particle at position $R$ we basically convolute the inhomogeneity with the spatial derivative of the solution profile in the absence of an inhomogeneity. This method has been successfully applied  in the case of LSs in the delayed Swift-Hohenbrg equation~\cite{tabbert17}. For a detailed derivation we refer the reader to \cite{tabbert17}. The potential reads:
\begin{align}
C\partial_R V(R)=\int \left\{\text{Re}[\partial_\xi E_\text{hs}(\xi)]~ Ae^{-(\xi+R)^2/B}\right\}d\xi,
\label{pot}
\end{align}
where $E_\text{hs}(\xi)$ refers to the stationary LS in the homogeneous case ($A=0$) and the dissipative constant $C=\int \partial_\xi \textbf{E}_\text{hs}(\xi)\cdot \partial_\xi \textbf{E}_\text{inhs}(\xi) d\xi$. In this case $\textbf{E}_\text{inhs}$ is the stationary LS solution on the center of the inhomogeneity (stable or unstable) written as a vector-function with the real and imaginary part as separate components. $\textbf{E}_\text{hs}(\xi)$ again is the solution without inhomogeneity, also written in vector form. The potentials for the two solutions shown as insets in Fig. \ref{fig:theta3} are depicted in Fig. \ref{fig:pot} for $\theta=3$, a constant inhomogeneity of $A=0.1$ and two different  values of $E_i=1.6$ (left) and $E_i=2.0$ (right). The potential model not only qualitatively describes the transition from an attracting to a repelling potential, it also provides numerically exact predictions of the position of stable solutions (orange lines). This may seem trivial in the case of an attracting inhomogeneity, since the stable solution in the center is explicitly regarded in Eq. \eqref{pot}. However, in the case of a repelling inhomogeneity pinning solutions on the side, the model also proves useful for a prediction of the position, although the calculation of the potential is only based on the unstable solution on the center $\textbf{E}_\text{inhs}$ and the solution without inhomogeneity $\textbf{E}_\text{hs}(\xi)$. The potential well model defined by Eq. \eqref{pot} therefore provides an easy way to estimate the effect of an inhomogeneity and the position of resulting stable solutions in the LLE, however it is restricted to the limit of small values of $A$, i.e. it does only reproduce the transition from an attracting to a repelling inhomogeneity at $A=0$ in a transcritical bifurcation and not the transition at finite values of $A$ in a pitchfork bifurcation. However, it still provides a good understanding of the effect of small inhomogeneities and therefore it seems promising to apply this method to other systems.

\section{Exploration of parameter space}
\begin{figure}
\hspace*{-0.35cm}
\includegraphics[width=0.5\textwidth]{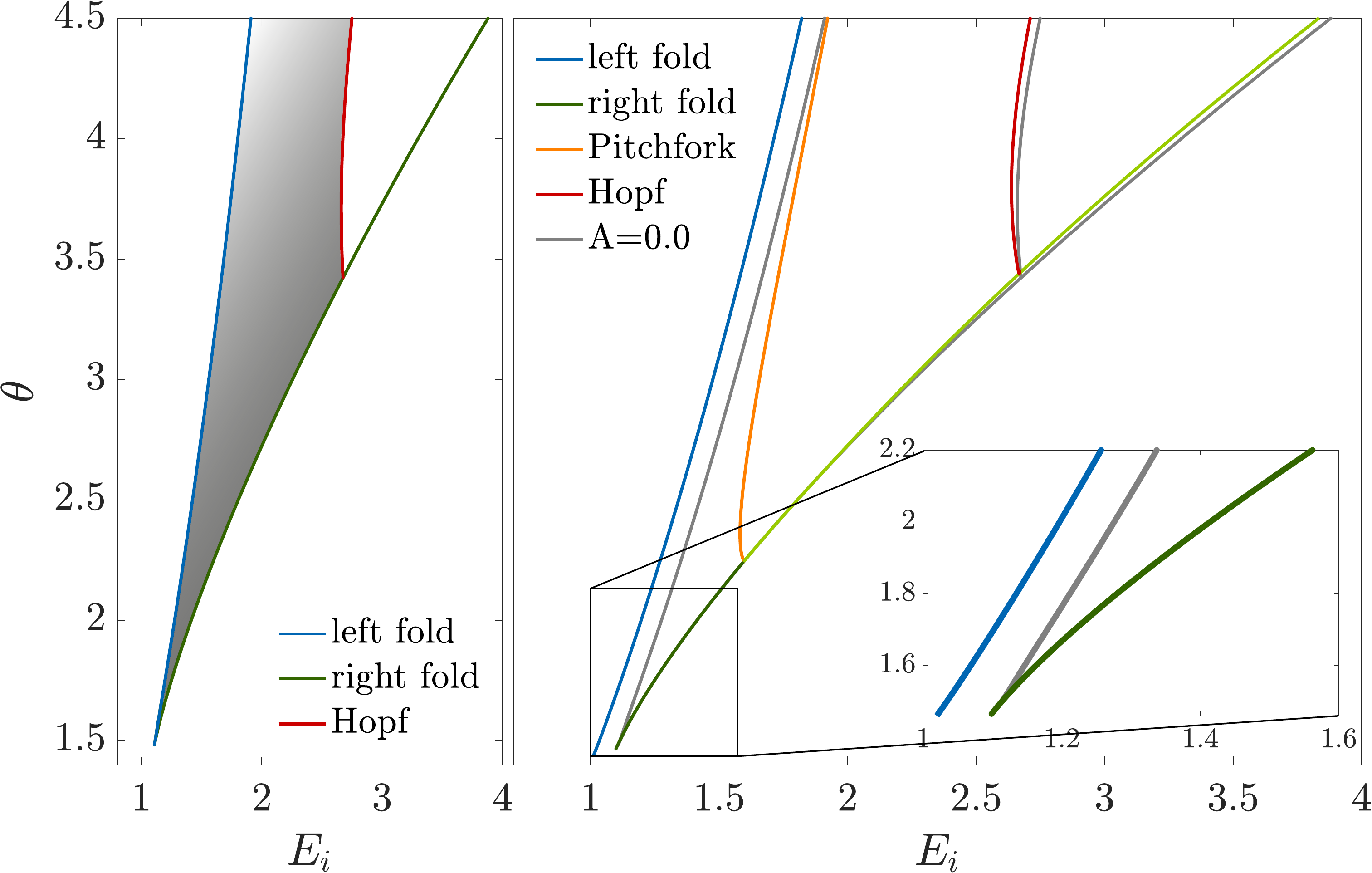}
 \caption{Left: Stability domain of a single LS (grey) in the ($E_i$, $\theta$) plane without an inhomogeneity. The stability regime is delimited by the position of the left (blue) and the right (green) fold of the homoclinic snaking diagram. For larger $\theta$, an Andronov-Hopf bifurcation (red) sets in, in which localized solutions loose their stability and start to oscillate. Right: Same diagram in the case of an inhomogeneity $A=0.1$. The results for $A=0.0$ are depicted in the background (grey). The left fold (blue) moves drastically to lower values of $E_i$ compared to the case $A=0$. For low values of $\theta$, this effect is relatively large compared to the overall region of stability which is depicted in the inset. The position of the right fold (green) is hardly affected by the inhomogeneity. The orange line marks the position of the pitchfork bifurcation, inducing a transition from stable solutions pinned on the center to solution pinned on the side of the inhomogeneity. I.e. on the right side of the orange line, solutions pinned on the side are stable up to the light green line. The position of the Andronov-Hopf bifurcation is hardly affected by the inhomogeneity.}
 \label{fig:param}
\end{figure}

So far, we have discussed the behavior of localized solutions in the LLE with and without inhomogeneity for fixed values of the detuning $\theta$. As suggested in Fig. \ref{fig:theta1.7} (bottom) and Fig. \ref{fig:theta3} (right panel) it is also possible to deploy numerical continuation techniques to track bifurcations or fold points in parameter space. Instead of altering a single continuation parameter (e.g. $E_i$ in the above examples) and approximating solutions of Eq. \eqref{LLE}, we now need an additional condition that defines the bifurcation (or fold) point. We then use $\theta$ as a primary continuation parameter and $E_i$ as a free parameter that is chosen accordingly to fulfill the aforementioned auxiliary condition. Results of this approach are shown in Fig. \ref{fig:param}. On the left panel of Fig. \ref{fig:param} fold and bifurcation point continuations have been performed for the case  $A=0$. Similar results obtained by means of numerical linear stability analysis have been obtained in \cite{Parra_Rivas_PRA_2014}. By following the left (blue) and right (green) fold of the single LS solution we can determine the region of stability for $\theta=2$. At larger values of the detuning, an Andronov-Hopf bifurcation leads to oscillations of LSs which result in a modulation of optical combs and is therefore undesirable. Tracking this bifurcation point in the $\theta$-$E_i$ space (red) provides the complete systematic description of the parameter space (grey shaded region) in which a stable single LS exists in the LLE. Figure \ref{fig:param} (right panel) provides the same analysis for the case of a small inhomogeneity $A=0.1$. The results for $A=0$ are depicted in grey there to provide a comparison. As shown in Fig. \ref{fig:theta1.7}, even for small values of $\theta$, the left fold point (blue) marking the onset of stability of localized solutions shifts drastically to smaller values of $E_i$, thus increasing the region of stable localized solutions. The shift remains approximately the same for all values of detuning. Therefore the relative growth of the region of stability due to the inhomogeneity is largest for small values of detuning in the snaking regime of the LLE (see inset). As already mentioned, the position of the right fold (dark and light green lines) remains almost unaltered by the inhomogeneity. The transition from a solution pinned on the center to a solution pinned on the side in a pitchfork bifurcation was already depicted for the case of $\theta=3$ in Fig. \ref{fig:theta3}. The orange line now marks this bifurcation position in the parameter space, i.e. solutions pinned on the center (side) are stable on the left (right) of the orange line. Note that the position of this line marks the position where the overall injection at the center reaches the ideal value as described in \cite{HendryPRA2018}. Finally, the onset of the Andronov-Hopf instability is hardly altered by the inhomogeneity (red line).

\section{Summary}
To summarize, the influence of the inhomogeneities on the stability of LSs in a mean-field LLE model for fiber resonators was studied. We have shown that the inhomogeneities not only allow for the control of the position of LSs, but also alter strongly their stability and bifurcation properties. We have constructed the bifurcation diagram associated with decaying oscillatory tails and showed that the stability regime of LSs significantly widens. Furthermore, in the parameter regime where the homoclinic snaking structure is lost, the effect of a given inhomogeneity is more complex, acting either attracting or repelling. To analyze the effect of the inhomogeneity and to calculate the position of stable solutions without much computational effort, we proposed to treat LSs in the vicinity of an inhomogeneity as an overdamped particle in a potential well. Finally, we have provided a full description of the stability region of LSs  in the LLE with and without inhomogeneity in terms of both detuning and injection.
\\\\
\section*{Acknowledgements}
F.T. received funds from the German Scholarship Foundation and the Center for Nonlinear Science
Münster.  M.T.  received  support  from  the  Fonds  National  de  la  Recherche  Scientifique  (Belgium). K.P. acknowledges the FWO Vlaanderen project G0E5819N  and the Methusalem foundation. 

\bibliographystyle{apsrev4-1}
\end{document}